\documentclass[twocolumn,prb,showpacs]{revtex4}
\usepackage{graphicx}
\usepackage{dcolumn}
\usepackage{bm}
\usepackage{amsmath}
\begin{document}
\title{Electromagnetic spin polarization on the surface of topological insulator}
\author{Tetsuro Misawa, Takehito Yokoyama, Shuichi Murakami}
\affiliation{Department of Physics, Tokyo Institute of Technology, Ookayama, Meguro-ku, Tokyo 152-8551, Japan}
\date{\today}
\begin{abstract}
We study the spin polarization of the electrons on the surface of topological insulators under a dc electric field or a circularly polarized light by using Keldysh Green's function formalism.
When a dc electric field $E_x \sim 10^{3} {\rm V/m}$ is applied, a spin polarization $\langle \sigma_{y} \rangle \simeq 5.2 \times 10^{-8}{\rm \AA^{-2}}$ is induced. Furthermore, we also find that a light illumination induces the out-of-plane component of the spin polarization as a result of the inverse Faraday effect. The magnitude of the spin polarization is proportional to the square of the lifetime $\tau$ and $\langle \sigma_{z}\rangle \simeq 2 \times 10^{-10}{\rm \AA^{-2}}$ for typical parameters. Finally, we investigate the spin polarization in the presence of the warping term. By the symmetry consideration of the system, we find that the out-of-plane spin polarization is cubic of the current and that the magnitude of the induced spin polarization depends on the direction of the applied current. 

\end{abstract}
\maketitle

\section{INTRODUCTION}
The three-dimensional topological insulator (TI) is an intriguing material, which is insulating in its bulk but has conducting surface states topologically protected by the time-reversal symmetry.~\cite{Hasan,Moore,Fu,Teo,Hsieh,Nishide}
The surface state of the TI forms Dirac cone band, and recent spin and angular resolved photoemission spectroscopy (spin-ARPES) experiments have revealed the Dirac band structure.
Also, other experimental reports on the TI include the observation of a massive Dirac fermion in magnetically doped TIs.~\cite{Chen2} 
Up to now, several materials, such as ${\rm Bi_{2}Te_{3}}$ and ${\rm Bi_{2}Se_{3}}$ have been identified as TIs.~\cite{Hsieh2,Xia,Sato,Kuroda}  

There are several theoretical predictions on phenomena unique to the TI.~\cite{Hasan,Qi,Qi2} 
Among them, it is worthy of attention that the spin and the momentum of an electron are locked to each other, which causes external-field-driven spin-related responses on the surface of the TI. This effect will not only feature  spintronics applications~\cite{Liu,Yokoyama1,Yokoyama2,Mondal,Raghu,Garate,Garate2,Yokoyama3,Burkov} but also bring up the possibilities of a new probe to surface states of TIs.
From this point of view, it is worthwhile to focus on the response of the surface states of TIs to external electric fields.

The spin-related response of the TI to electric fields is attributed to the spin-orbit coupling (SOC). A similar effect is also present in the Rashba system.~\cite{Rashba} The Rashba system has the Hamiltonian of the form $\frac{\bm{k}^{2}}{2m}+\alpha (\bm{\sigma}\times \bm{k})_{z}$, where $\sigma_{i}$'s are the Pauli's matrices in spin space and $\alpha$ is a constant unique to the material. Since the spins of the electrons in the Rashba system are aligned in the clockwise or counterclockwise manner in the $\bm{k}$-space, it is seen that dc currents induce spin accumulations.~\cite{Ede90,Ino03,Sinova}
Because of this fact, we expect that interesting spin-related phenomena also appear on the surfaces of the TI. 

Magneto-optical Kerr and Faraday effects of a TI thin film have been studied in Refs.~[\onlinecite{Tse,Maciejko}].
The inverse Faraday effect (IFE) is an effect that circularly polarized light induces the stationary magnetization.~\cite{Pitaevskii,van der Ziel,Pershan,Landau,Ede98,Hertel,Woodford}
Recent experimental progress has made it possible to reverse magnetization in magnets with a circularly polarized light through the IFE,~\cite{Kimel,Bigot,Kirilyuk} and now the IFE is a powerful method of ultrafast magnetization manipulation.
In view of this, it is quite timely to study the IFE of the TI. 

Additionally, it has been clarified that the effect of the warping term may be remarkable in ${\rm Bi_{2}Te_{3}}$.~\cite{Chen}
The warping term arises accordingly to the space group of the crystal.
Theoretically, based on the $k\cdot p$ theory it is found that the warping is due to the cubic term of $k$ in the Hamiltonian.~\cite{Fu09,Lee09} It is easily seen that on the warped band structure, the spins of the eigenstates have out-of-plane components. Therefore it is naively expected that dc current flowing through the surface will induce out-of-plane spin polarizations. 

In this paper,
we study the spin polarization of the electrons on the surface of TIs under a dc electric field or a circularly polarized light by using Keldysh Green's function formalism.
When a dc electric field $E_x \sim 10^{3} {\rm V/m}$ is applied, a spin polarization $\langle \sigma_{y} \rangle \simeq 5.2 \times 10^{-8}{\rm \AA^{-2}}$ is induced. Furthermore, we find that  a light illumination induces the out-of-plane component of the spin polarization as a result of the IFE. The magnitude of the spin polarization is proportional to the square of the lifetime $\tau$ and $\langle \sigma_{z}\rangle \simeq 2 \times 10^{-10}{\rm \AA^{-2}}$ for typical parameters. Finally, we also investigate the spin polarization in the presence of the warping term. By the symmetry consideration of the system, we find that the out-of-plane spin polarization is cubic of the current and that the magnitude of the induced spin polarization depends on the direction of the applied current.

The organization of this paper as follows. In section II, we explain the formalism. In section III, we present the calculated results of the current-induced spin polarization and the spin polarization by circularly polarized light, and discuss the spin polarization in the presence of the warping term.
In section VI, we summarize our results. In Appendices A and B, details of the calculation of Green's functions are presented. In Appendix C, we compare our results with those in the Rashba system and  in Appendix D, calculation based on the Kubo formula is presented.

\section{FORMULATION}
We consider a surface of a TI along the $xy$ plane. The Hamiltonian of the surface state is given by 
\begin{align}
H_{0} = v( \bm{\sigma} \times \bm{k})_{z} =
	\left(
		\begin{array}{cc}
			0 & ivk_{-} \\
			-ivk_{+}& 0
		\end{array}
	\right) ,
\label{ti_model_Hamiltonian}
\end{align}
where $k_{\pm} = k_{x}\pm i k_{y}$, $\sigma_{i}$'s are the Pauli matrices, $v$ is a constant representing the velocity, and we choose the unit so that $\hbar = 1$.
The eigenvalues are $E_{\bm{k}s} = s v k$ with $s = \pm $ and $k=\sqrt{k_{x}^{2}+k_{y}^{2}}$, and the corresponding eigenstates are $\phi_{\bm{k},s}(\bm{r})=e^{i\bm{k}\cdot \bm{r}}/{\sqrt{2L^{2}}}(1,\, -isk_{+}/k)$ with $L^{2}$ being the size of the system.

In what follows, we introduce the Keldysh Green's function formalism, which allows us to calculate physical quantities in non-equillibrium states. At first, we define the lesser Green's function with the field operators $c$ and $c^{\dagger}$ as $G^{<}(\bm{r},t,\bm{r}',t') \equiv i\langle c_{\rm H}^{\dagger}(\bm{r}',t')c_{\rm H}(\bm{r},t) \rangle$, which takes a 2 $\times$ 2 matrix form since these field operators are spinors.
The $c^{(\dagger)}_{\rm H}$ is the Heisenberg representation of the annihilation (creation) operator. The brackets $\langle \cdots \rangle$ mean both quantum and thermal averages. 

Usually, it is impossible to calculate the Green's function exactly for complicated Hamiltonian, and therefore the pertubative method for the Green's function is employed. 
We now consider the entire Hamiltonian written in the form $H = H_{0}+V$, where the $H_{0}$ is the unperturbative Hamiltonian, which has been solved exactly, and the $V$ is a perturbative term.
The calculation is done with the Dyson's equation. With $g_{0}$ and $g$ being the unperturbed and perturbed Green's functions, respectively, the Dyson's equation yields 
\begin{align}
g^{<}=g_{0}^{<} + g_{0}^{\rm r}Vg^{<}+g_{0}^{<}Vg^{\rm a}. \label{shaei_koushiki}
\end{align}
$g_{0}^{\rm r}$ and $g_{0}^{\rm a}$ are the retarded and advanced Green's functions, which are expressed as $g_{0}^{\rm r} = (\omega - {H_{0}} + i0)^{-1}$ and $\ g_{0}^{\rm a} = (\omega - {H_{0}} - i0)^{-1}$ respectively. 
We utilize relations between these Green's functions $g_{0}^{<}$, $g_{0}^{\rm r}$ and $g_{0}^{\rm a}$ : $g_{0\bm{k},\omega}^{<} = f(\omega ) (g_{0\bm{k},\omega }^{\rm a}-g_{0\bm{k},\omega }^{\rm r}) \label{f(g-g)}$ with $f(\omega)$ being the Fermi distribution function.~\cite{Haug}
Then we can use Eq.~(\ref{shaei_koushiki}) iteratively to obtain arbitrary higher order perturbations as $g^{<} = g_{0}^{<} + g_{0}^{\rm r}V g_{0} ^{<}+g_{0}^{<}Vg_{0}^{\rm a} +  g^{\rm r}_0 Vg^{\rm r}_0 Vg^{<}_0 +g^{\rm r}_0 Vg^{<}_0 V g^{\rm a}_0 +g^{<}_0 V g^{\rm a}_0 Vg^{\rm a}_0  + \cdots$.

\begin{figure}[htbp]
	\begin{center}
		\includegraphics[width=8cm, keepaspectratio, clip]{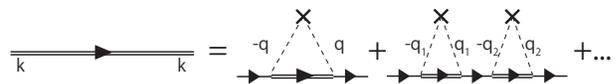}
		\caption{The perturbed Green's function obtained by the self-consistent Born approximation. The single and double lines mean the unperturbed and perturbed Green's functions, respectively, and the dashed lines are scattering potentials by the impurities.}
		\label{fig:born_approx}
	\end{center}
\end{figure}

Here, we consider spin-independent impurities $V=\sum_{i=1}^{N_{\rm imp}} V_{\rm imp} \delta (\bm{r}-\bm{r}_{i})$, where $N_{\rm imp}$ is the number of the impurities in the system. Supposing that the impurities are dilute, the self-consistent equation can be solved perturbatively. We take the random average over all possible configulations of the impurities to recover the translation symmetry of the system. The self-consistent Born approximation is equivalent to the Born approximation in the limit of dilute impurities.  We assume this limit in the following, and then we get the self-energy as
\begin{align}
\Sigma^{\rm r(a)} & = \frac{n_{\rm imp}V_{\rm imp}^{2}}{L^{2}}\sum_{\bm{k}}g_{\bm{k},\omega}^{\rm r(a)} \simeq \frac{n_{\rm imp}V_{\rm imp}^{2}}{L^{2}} \sum_{\bm{k}} g^{\rm r(a)}_{0\bm{k},\omega} \nonumber \\
& \simeq \mp i \frac{n_{\rm imp}V_{\rm imp}^{2}}{4 v^{2}}|\omega|\sigma_{0}, \label{self_energy}
\end{align}
where $\sigma_{0}$ is the identity matrix in spin space and $n_{\rm imp}=N_{\rm imp}/L^{2}$ is the concentration of the impurities. We here supposed $|\epsilon_F|\tau \gg 1$ and the real part of the self-energy can be neglected. Using the self-energy, we obtain the modified Green's functions: $g^{\rm r(a)} = (\omega -{H_0} -\Sigma ^{\rm r(a)})^{-1}$.
We define $\eta_{\omega} = n_{\rm imp}V_{\rm imp}^{2}|\omega|/4 v^{2}$, and the relaxation time is then given by $\tau = 1/2\eta_{\epsilon_{\rm F}}$. In the following sections, we assume that $|\epsilon_{\rm F}|\tau \gg 1$ and $\Omega \tau \ll 1$, where $\Omega$ is the frequency of the external field. The explicit form of the modified Green's function reads
\begin{align}
g_{\bm{k},\omega}^{\rm r(a)} = \frac{1}{(\omega \pm i \eta_{\omega})^{2}-v^2 k^2 }\left(\begin{array}{cc}\omega & -ivk_{-}\\ivk_{+}&\omega \end{array}\right),
\end{align}
which corresponds to the diagram in Fig.~\ref{fig:born_approx}.

\section{RESULTS}

\subsection{Current-induced spin polarization}
In this section, we will study the spin polarization in response to the dc electric field applied parallel to the surface. 
When we consider the dc external field, it makes the calculation easier to introduce the field oscillating with frequency $\Omega$ and then take the limit $\Omega \searrow +0$. In the following, we consider the spatially uniform field so that the wave vector $\bm{k} $ becomes a good quantum number. We omit the wave number $\bm{k}$ of quantities for brevity.

The external field is expressed by  the vector potential $\bm{A}(t)$ which satisfies $\bm{E}(t) = - \dot{\bm{A}}(t)$, and then the perturbation term reads $V(t)=-\bm{j}\cdot \bm{A}(t)$, where $\bm{j}$ is the current operator. The current operator is given by $j_{x} = -e\dot{x} = ev\sigma_{y}$, $j_{y} = -e\dot{y} = -ev\sigma_{x}$. When the electric field is along the $+x$ direction, the perturbation term yields $V(t)=evE_{x}e^{i\Omega t}\sigma_{y}/(i\Omega)$, whose Fourier transform is $2\pi \delta (\Omega-\Omega') \tilde{V}_{\Omega'}= 2\pi \delta (\Omega-\Omega') evE_{x}\sigma_{y}/(i\Omega')$. Using Eq.~(\ref{shaei_koushiki}), we obtain up to linear order in $V$;
\begin{widetext}
\begin{align}
G^{<}_{\omega, \omega + \Omega'} & \simeq 2\pi \delta(\Omega') g_{\omega}^{<}+2\pi \delta (\Omega-\Omega')
 \left[ f(\omega + \Omega')g^{\rm r}_{\omega} \tilde{V}_{\Omega'}(g^{\rm a}_{\omega + \Omega'}-g^{\rm r}_{\omega + \Omega'})
 + f(\omega) (g^{\rm a}_{\omega}-g^{\rm r}_{\omega})\tilde{V}_{\Omega'}g^{\rm a}_{\omega + \Omega'} \right] \label{first_order_perturbation}, \nonumber \\
& = 2\pi \Biggl[ \delta(\Omega') g_{\omega}^{<} +\delta(\Omega-\Omega')
	\biggl\{ -f(\omega) \left( g_{\omega}^{\rm r}\tilde{V}_{\Omega'}g_{\omega}^{\rm r}-({\rm r}\leftrightarrow {\rm a}) \right) \biggr. \Biggr.
+ \nonumber \\
& \hspace{13em} \Biggl. \biggl. \Omega' \left\{ f(\omega) \left(g_{\omega}^{r}\tilde{V}_{\Omega'} (g_{\omega}^{\rm r})^{2} - ({\rm r}\leftrightarrow {\rm a}) \right)+ f'(\omega) g_{\omega}^{\rm r}\tilde{V}_{\Omega'}(g_{\omega}^{\rm a}-g_{\omega}^{\rm r})
	\right\} \biggr\}
	\Biggr],
\end{align}
\end{widetext}
where $f(\omega)=(e^{(\omega-\epsilon_{\rm F})/k_{\rm B}T}+1)^{-1}$ is the Fermi distribution function. We here expanded the Fermi distribution function and Green's functions in terms of $\Omega$.
In the above expansion, we can show that the $g_{\omega}^{\rm r(a)}\tilde{V}_{\Omega'}g_{\omega}^{\rm r(a)}$ terms vanish by the integration in $k$-space (See Appendix A). Furthermore, the $g_{\omega}^{\rm r(a)}\tilde{V}_{\Omega}(g_{\omega}^{\rm r(a)})^{2}$ terms do not contribute to physical quantities. The higher order derivatives of Green's functions give higher order terms of $\Omega$ and can be neglected. Then, only the $g_{\omega}^{\rm r}\tilde{V}_{\Omega'}g_{\omega}^{\rm a}$ term contributes to the result. Therefore, the Keldysh formalism within the linear response is apparently equivalent to the Kubo formula approach. Calculation based on the Kubo formula will be also discussed in the Appendix D.

\begin{figure}[htbp]
		\begin{center}
		\includegraphics[width=7.2cm, keepaspectratio, clip]{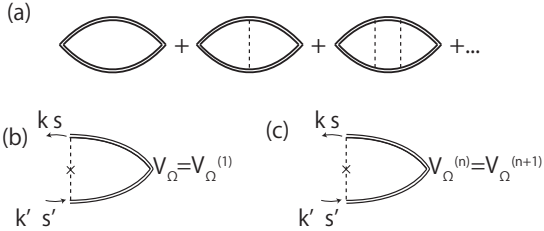}
		\caption{We have to include the diagrams shown in (a) to satisfy the Ward-Takahashi identity, which ensures the gauge invariance. (b) and (c) are the iterative method we use to get the n-th ladder diagrams shown above.}
		\label{fig:ladder}
	\end{center}
\end{figure}

Next, we calculate the sum of the ladder diagrams in Fig.~\ref{fig:ladder}, which corresponds to the vertex correction.

To calculate these ladder terms, we use the iterative procedure. First, we consider the diagram with impurity scattering in Fig.~\ref{fig:ladder}(b). The corresponding term is $ \langle V_{\rm imp} g_{\omega}^{\rm r} V_{\Omega}g_{\omega}^{\rm a}\tilde{V}_{\rm imp}\rangle_{\rm AV}$, 
where the $\langle \cdots \rangle_{\rm AV}$ is the random average over the impurity configurations. Using the concrete expressions of Green's functions and perturbation terms, we obtain
$\langle \tilde{V}_{\rm imp}g_{\omega}^{\rm r}\tilde{V}_{\Omega}^{} g_{\omega}^{\rm a}\tilde{V}_{\rm imp}\rangle_{\rm AV}\equiv V_{\rm \Omega}^{(1)} = \tilde{V}_{\Omega}/{2}$. This calculation shows that the first-order correction for the vertex function becomes half the original one. Similarly, the higher correction terms yield $\langle \tilde{V}_{\rm imp}g_{\omega}^{\rm r} \tilde{V}_{\Omega}^{(n)} g_{\omega}^{\rm a}\tilde{V}_{\rm imp}\rangle_{\rm AV} \equiv \tilde{V}_{\Omega}^{(n+1)} =\tilde{V}_{\Omega}^{(n)}/{2}$, therefore $\tilde{V}_{\Omega}^{(n)}= \tilde{V}_{\Omega}/2^{n}$. Summing up all the ladder diagrams, we obtain $\sum_{n=1}^{\infty} \tilde{V}_{\Omega}^{(n)}=\sum \tilde{V}_{\Omega}/2^{n}=\tilde{V}_{\Omega}$. This means that when the ladder terms are considered, the original $V_{\Omega}$ should be doubled  $\hat{V}_{\Omega}\equiv \tilde{V}_{\Omega}+\sum_{n=1}^{\infty}\tilde{V}_{\Omega}^{(n)} = 2\tilde{V}_{\Omega}$. After substituting $\tilde{V}_{\Omega}$ with $\hat{V}_{\Omega}$ and the Fourier transform, we obtain the perturbed lesser Green's function including the vertex correction at zero temperature of the form;
\begin{align}
G^{<}(\bm{r}, t, \bm{r}, t) & \simeq -\frac{1}{L^2}\sum_{\bm{k}}\int \frac{\mathrm{d}\omega}{2\pi}\Omega f'(\omega) g_{\omega}^{\rm r}\hat{V}_{\Omega} g^{\rm a}_{\omega} \nonumber \\
& = \frac{ievE_{x}}{2\pi n_{\rm imp}V_{\rm imp}^{2}}\sigma_{y}.
\end{align}

The local spin density of the spin polarization $\langle \sigma_{i} \rangle$ at the position $\bm{r}$ and time $t$ is given by making the trace of the product of the Pauli's matrix and the lesser Green's function: 
\begin{align}
\langle \sigma_{i} \rangle = -i {\rm Tr}\, \sigma_{i} G^{<}(\bm{r},\, t,\bm{r},\, t)
\end{align}
where $i=x,y,z$. Therefore, the expectation value of $\sigma_{y}$ reads
\begin{align}
\langle {\sigma}_{y} \rangle = -i{\rm Tr}\, \sigma_{y} \left[ \frac{ievE_{x}}{2\pi n_{\rm imp}V_{\rm imp}^{2}}\sigma_{y}\right] = \frac{evE_{x}}{\pi n_{\rm imp}V_{\rm imp}^{2}}.
\end{align}
In a similar way, the expectation values of $\sigma_{x}$ and $\sigma_{z}$ are calculated  to be zero.
As for the charge conductivity, it can be calculated in a similar way:
\begin{align}
\sigma_{xx} = \frac{e^{2}v^{2}}{\pi n_{\rm imp}V_{\rm imp}^{2}}. \label{keldysh_conductivity}
\end{align}

Note that the conductivity is independent of the Fermi energy within the self-consistent Born approximation, with the real part of the self-energy neglected. The form of the conductivity (Eq.~(\ref{keldysh_conductivity})) is the same as that of the graphene for the short-ranged impurities,~\cite{Sho98} which also has a similar linear dispersion. This relation can be understood in an intuitive way by noting a difference of the degrees of freedom: Graphene has two Dirac cones at K and K' points in its Brillouin zone, and each Dirac cone is spin-degenerate, and hence is two-fold.
Therefore, the number of freedom of the graphene that contributes to the conductivity is four times greater than that of the TI. However, the contribution from  non-magnetic impurities in graphene is four times larger than that in the TI because of the spin-momentum locking. As a result, these factors cancel out each other and we get the same expression of the conductivities.

Experimentally, the presence of the bulk conduction carrier would hinder precise estimations of longitudinal conductivity of surfaces of TIs. The bulk conduction is mainly due to vacancies of Se and Te atoms, ionic impurities or lattice defects.
However, the spin polarization discussed above is a surface effect, and its measurement give us the physics of helical surface states separated from the bulk. In fact, there is a relation between the spin response function $\phi_{y}$ and the longitudinal conductivity $\sigma_{xx}$ as $\sigma_{xx} = ev\phi_{y}$, which offers a possibility of a new probe to the surface states of TIs.

Now, we make a realistic estimate of the spin polarization. We here suppose that it is a good approximation that the conductivity on surface of TI $\sigma_{xx}$ is as large as that of graphene; $\sigma_{xx}^{\rm graphene} = 2.5 \times 10^{-4} {\rm \Omega^{-1}}$. The parameter $v$ in the model Hamiltonian (Eq.~(\ref{ti_model_Hamiltonian})) is $v \simeq 2.0 {\rm eV \AA}$ from the observation of surface states of ${\rm Bi_{2}Se_{3}}$ by the ARPES experiment,~\cite{Chen} which is typically larger than the Fermi velocity in graphene.
We assume applying the electric field of strength $10^{3} {\rm V/m}$, and then the induced spin polarization is calculated to be $\langle \sigma_{y} \rangle \simeq 5.2 \times 10^{-8}{\rm \AA^{-2}}$. The corresponding magnetization can be obtained by multiplying it by the Bohr magneton $\mu_{\rm B}$ and the electron spin g-factor.

\subsection{Inverse Faraday effect}
\begin{figure}[htbp]
\begin{center}
\includegraphics[width=7cm, keepaspectratio, clip]{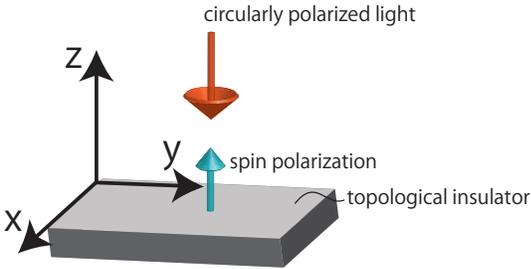}
\caption{(color online) The circularly polarized light propagating along the $-z$ axis induces the spin polarization $\langle \sigma_{z}\rangle$.}
\label{fig:ife_setup}
\end{center}
\end{figure}

In this section, we investigate the nonlinear effect caused by a circularly polarized  light, namely the IFE, where a dc spin polarization is induced by a light illumination. The axes are set as in Fig.~\ref{fig:ife_setup}.

We employ the gauge to set the scalar potential to be zero and hence $\dot{\bm{A}}(t)=-\bm{E}(t)$. Then, the perturbation term is $V(t) = -\bm{j}\cdot \bm{A}(t)+{\rm h.c.}$
We consider spatially uniform electric field $\bm{E}(t) = \bm{E}_{\Omega} e^{i\Omega t}$ with $\bm{E}_{\Omega} = (E_{\Omega}^{(x)},\, E_{\Omega}^{(y)},\, 0)$, then Fourier transform of the perturbation Hamiltonian is $V_{\bm{k}, \Omega'}  = 2\pi L^{2} \delta_{\bm{k},0} [ \delta (\Omega'-\Omega) \tilde{V}_{\Omega} + \delta(\Omega'+\Omega) \tilde{V}_{\Omega}^{\dagger} ]$, where $\tilde{V}_{\Omega} = -\bm{j}\cdot \bm{E}_{\Omega}/i\Omega$ and $\tilde{V}_{\Omega}^{\dagger}$ is the hermitian conjugate of $\tilde{V}_{\Omega}$.
When the polarization is left circularly polarized, the electric field is $\bm{E}_{\Omega} = {\cal E}(1,\, -i,\, 0)$ with ${\cal E}$ being real, and $V_{\Omega}$ and $V_{\Omega}^{\dagger}$ are reduced to be
\begin{align}
\tilde{V}_{\Omega} = \frac{2ev}{\Omega}
\left(
	\begin{array}{cc}
		0 & 0 \\
		{\cal E} & 0
	\end{array}
\right) ,\ \ 
\tilde{V}_{\Omega}^{\dagger} = \frac{2ev}{\Omega} \left(
\begin{array}{cc}
	0 & {\cal E} \\
	0 & 0 \\
\end{array}
\right) \label{second_perturbation}.
\end{align}

\begin{figure}[htbp]
	\begin{center}
		\includegraphics[width=7.2cm, keepaspectratio, clip]{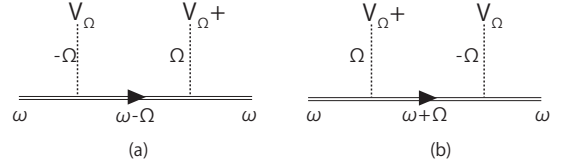}
		\caption{ (a) and (b) correspond to the terms $g_{\omega}Vg_{\omega -\Omega}Vg_{\omega}$ and $g_{\omega}Vg_{\omega +\Omega}Vg_{\omega}$ respectively. These terms give the diagonal components in the Fourier transformed $\omega$ space.}
		\label{fig:second_diagram}
	\end{center}
\end{figure}

Because we are especially interested in the dc response of the TI to the oscillating field, we calculate the second order perturbation with the Keldysh formalism, and take into account terms in Fig.~\ref{fig:second_diagram}.

By using the Dyson's equation iteratively, we obtain the expression up to the second order perturbation with respect to $V$:
\begin{align}
G^{<} & \simeq  g^{<}+g^{\rm r}Vg^{<}+g^{<}Vg^{\rm a} \nonumber \\
&+g^{\rm r}Vg^{\rm r}Vg^{<}+g^{\rm r}Vg^{<}Vg^{\rm a}+g^{<}Vg^{\rm a}Vg^{\rm a}.
\end{align}

In particular, the diagonal elements in $\omega$ are necessary to calculate the dc response of the system. $G^{<}_{\omega}$ is given by 
\begin{align}
G^{<}_{\omega}  \simeq  g_{\omega}^{<}
	& + f(\omega) \left\{ \left( g_{\omega}^{\rm r}\tilde{V}_{\Omega}g^{\rm r}_{\omega + \Omega}\tilde{V}_{\Omega}^{\dagger}(g_{\omega}^{\rm a}-g_{\omega}^{\rm r}) \right. \right.  \nonumber \\
	 & \hspace{3.8em} \left. \left. + g_{\omega}^{\rm r}\tilde{V}_{\Omega}^{\dagger}g^{\rm r}_{\omega - \Omega}\tilde{V}_{\Omega}(g_{\omega}^{\rm a}-g_{\omega}^{\rm r}) \right) -{\rm h.c.} \right\} \nonumber \\
	 & + f(\omega + \Omega)g_{\omega}^{\rm r}\tilde{V}_{\Omega}(g_{\omega + \Omega}^{\rm a}-g_{\omega + \Omega}^{\rm r})\tilde{V}_{\Omega}^{\dagger}g_{\omega}^{\rm a} \nonumber \\
	 & + f(\omega - \Omega)g_{\omega}^{\rm r}\tilde{V}_{\Omega}^{\dagger}(g_{\omega - \Omega}^{\rm a}-g_{\omega - \Omega}^{\rm r})\tilde{V}_{\Omega}g_{\omega}^{\rm a} \label{second_order_perturbation},
\end{align}
where we used the formula $g^{<}_{\omega} = f(\omega)(g_{\omega}^{\rm a}-g_{\omega}^{\rm r})$. Notably, the first perturbative terms, which include a single $\tilde{V}_{\Omega}$ , do not contribute to the result, because they do not contribute to the diagonal $\omega$ components. 

To proceed further, we assume that $\Omega$ is small and expand the Fermi distribution function and Green's functions in terms of $\Omega$. We expect that we have to expand at least up to the third-order with respect to $\Omega$ because the perturbation terms $\tilde{V}_{\Omega}$ and $\tilde{V}_{\Omega}^{\dagger}$ are inversely proportional to $\Omega$.
Actually, the terms up to second order perturbation in $\Omega$ become zero (See Appendix B). The leading terms in Eq.~(\ref{second_order_perturbation}) are therefore reduced to 
\begin{align}
	\Omega^{3} \biggl( & \frac{1}{3!}f(\omega) \left[ -g_{\omega}^{\rm r}\tilde{V}_{\Omega}\frac{\partial^{3}g_{\omega}^{\rm r}}{\partial \omega^{3}}\tilde{V}_{\Omega}^{\dagger}g_{\omega}^{\rm r}+g_{\omega}^{\rm r}\tilde{V}_{\Omega}^{\dagger} \frac{\partial^{3}g_{\omega}^{\rm r}}{\partial \omega^{3}}\tilde{V}_{\Omega}g_{\omega}^{\rm r} - {\rm h.c.}\right] \biggr. \nonumber \\
  + & \frac{1}{2!} f'(\omega) \left[ g_{\omega}^{\rm r}\tilde{V}_{\Omega}\frac{\partial^{2} g^{\rm a-r}_{\omega}}{\partial \omega^{2}} \tilde{V}_{\Omega}^{\dagger} g_{\omega}^{\rm a}-g_{\omega}^{\rm r}\tilde{V}_{\Omega}^{\dagger} \frac{\partial^{2} g_{\omega}^{\rm a-r}}{\partial \omega^{2}}\tilde{V}_{\Omega} g_{\omega}^{\rm a} \right] \nonumber \\
 +  & \frac{1}{2!} f''(\omega) \left[ g_{\omega}^{\rm r}\tilde{V}_{\Omega}\frac{\partial g^{\rm a-r}_{\omega}}{\partial \omega} \tilde{V}_{\Omega}^{\dagger} g_{\omega}^{\rm a}-g_{\omega}^{\rm r}\tilde{V}_{\Omega}^{\dagger} \frac{\partial g_{\omega}^{\rm a-r}}{\partial \omega}\tilde{V}_{\Omega}g_{\omega}^{\rm a} \right] \nonumber \\
 \biggl. + & \frac{1}{3!} f'''(\omega) \left[ g_{\omega}^{\rm r}\tilde{V}_{\Omega} g^{\rm a-r}_{\omega} \tilde{V}_{\Omega}^{\dagger} g_{\omega}^{\rm a}-g_{\omega}^{\rm r}\tilde{V}_{\Omega}^{\dagger} g_{\omega}^{\rm a-r}\tilde{V}_{\Omega}g_{\omega}^{\rm a}\right]\biggr),
\end{align}
where $g_{\omega}^{\rm a-r}= g_{\omega}^{\rm a}-g_{\omega}^{\rm r}$. Second and higher derivative terms of the Fermi distribution function is partially integrated, which yields only the first order derivative of the Fermi distribution function.
With the partial integration and the Fourier transformation, we get the perturbed Green's function $G^{<}(\bm{r},t,\bm{r},t)$. From this Green's function, we calculate the local spin density as
\begin{align}
\langle \sigma_{z} \rangle = -i {\rm Tr}[{\sigma}_{z} G^{<}(\bm{r},t,\bm{r},t)] = \frac{k_{\rm F}^{2}}{4\pi}K_{\Omega}, \label{ife_green_real}
\end{align}
where $k_{\rm F}^{2}/4\pi$ is the 2D electron density, and $K_{\Omega}$ is represented with $\gamma = \eta_{\epsilon_{\rm F}}/\epsilon_{\rm F}$ as
\begin{widetext}
\begin{align}
K_{\Omega} = \frac{-ie^{2}v^{2}(\bm{E}_{\Omega} \times \bm{E}_{\Omega}^{*})_{z}\Omega}{\epsilon_{\rm F}^{5}} \left[ \frac{2\gamma (3-\gamma^{2})}{9\pi (1+\gamma^{2})^{3}} + \frac{2\gamma (3+2\gamma^{2}+3\gamma^{4}) + 3(1-\gamma^{2} )(1+\gamma^{2})^{2}({\pi}{\rm sgn} (\gamma) - 2\arctan \gamma )}{12\pi \gamma^{2}(1+\gamma^{2})^{2}} \right].
\label{K_Omega}
\end{align}
\end{widetext}
$K_{\Omega}$ is the averaged spin polarization per electron. Note that $|\gamma |\sim |(\epsilon_{\rm F}\tau)^{-1}| \ll 1$. In the limit of $\gamma \to 0$, Eq.~(\ref{K_Omega}) gives
\begin{align}
K_{\Omega} \simeq {\rm sgn} (\gamma) \frac{-i e^{2} v^{2}(\bm{E}_{\Omega}\times \bm{E}_{\Omega}^{*})_{z} \tau^{2} \Omega}{\epsilon_{\rm F}^{3}},
\end{align}
which is proportional to the square of the relaxation time $\tau$. In this limit, the local spin density $\langle \sigma_{z}\rangle$ yields
\begin{align}
\langle \sigma_{z} \rangle & = {\rm sgn}(\epsilon_{\rm F}) \frac{-ie^{2}(\bm{E}_{\Omega}\times \bm{E}_{\Omega}^{*})_{z} \tau^{2}\Omega}{4\pi \epsilon_{\rm F}}.
\end{align}

Notice that a finite spin polarization per unit area, $\langle \sigma_{z} \rangle$, is generated as a result of the light illumination.
The spin polarization is proportional to $(\bm{E}_{\Omega}\times \bm{E}_{\Omega}^{*})_{z}$ and hence the IFE is dependent on the helicity of the applied electric field. It is seen that when the polarization of the incident light is reversed or equivalently we change the field from $\bm{E}_{\Omega}={\cal E}(1,\, -i,\, 0)$ to $\bm{E}_{\Omega}={\cal E}(1,\, i,\, 0)$, then the induced spin polarization reverses.
Apparently $\langle \sigma_{z}\rangle$ is zero for a linearly polarized light. Although the IFE also occurs under the illumination of elliptic-polarized lights, it is most efficient in the case of circular polarized lights. For example, if the external field is $\bm{E}_{\Omega} = {\cal E}(1,\, -i,\, 0)$, then $\bm{E}_{\Omega}^{*} = {\cal E}(1,\, +i,\, 0)$ yields $\bm{E}_{\Omega}\times \bm{E}_{\Omega}^{*} = 2i{\cal E}^{2} \bm{e}_{z}$ with $\bm{e}_{z}$ being the unit vector along the $z$ axis.
As for the frequency dependence, the spin polarization by the IFE is proportional to $\Omega$ and vanishes in the limit of $\Omega \searrow 0$.

The polarization is proportional to the square of the relaxation time $\tau$. This is in stark contrast to the case of the Rashba system, where it is proportional to $\tau^{-2}$ for high light frequency.~\cite{Ede98} The resulting spin polarization is independent of the sign of $\epsilon_F$. Therefore, for a left-circularly polarized light, we have $-i(\mathbf{E}_{\Omega}\times\mathbf{E}^{*}_{\Omega})_{z}>0$ and $\langle\sigma_z\rangle>0$. 

At the end of this section, let us estimate the spin polarization for the realistic parameters; $1/2 \tau \simeq 1{\rm meV}$, $\Omega \simeq 1{\rm meV}$, $\epsilon_{\rm F}\simeq 10^{2}{\rm meV}$ and $v\simeq 2{\rm eV\AA}$~\cite{Chen}. When the electric field is of strength ${\cal E} \simeq 1\times 10^{4}{\rm V/m}$, then $K_{\Omega} = 1 \times 10^{-6}$. For the typical surface state of the TI, the $2$D electron density is $2 \times 10^{-4}{\rm \AA}^{-2}$ then the spin polarization density is $\langle \sigma_{z} \rangle \simeq 2 \times 10^{-10}{\rm \AA^{-2}}$. As noted in the former section, we can translate this value to the magnitude of magnetization by multiplying it by Bohr magneton $\mu_{\rm B}$ and the electron spin g-factor.

\subsection{Warping effect}
\begin{figure}[htbp]
	\begin{center}
		\includegraphics[width=5cm, keepaspectratio, clip]{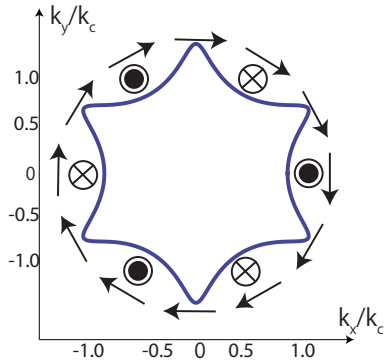}
		\caption{(color online) The spin configuration of the eigenvectors for the $s=+$ band. The arrows in this figure indicate the in-plane spin components of the eigenstates and the solid circle (cross) indicates the out-of-plane spin components. At the cusp points the spin lies completely in the $xy$ plane just as in the case without the warping term. However, at other points the spin has nonzero $\sigma_{z}$ component, which is the largest in its magnitude at the valley points.}
		\label{fig:warping_spins}
	\end{center}
\end{figure}

In this section, we investigate effects of the warping term. In terms of the crystal structure, ${\rm Bi_{2} Te_{3}}$ has the rhombohedral structure~\cite{Chen} and  its band structure has ${\rm C_{\it 3v}}$ symmetry. This effect have been treated theoretically in Refs.~[\onlinecite{Fu09,Lee09}].
According to these studies, the effective Hamiltonian includes the hexagonal warping term and takes the form
\begin{align}
{\cal H} = v(\bm{\sigma}\times \bm{k})_{z} +\frac{\alpha}{2}(k_{+}^{3}+k_{-}^{3})\sigma_{z}. \label{warping_hamiltonian}
\end{align}
The eigenstates and the eigenvalues are given by
\begin{align}
\psi_{+} = \frac{e^{i\bm{k} \cdot \bm{r}}}{\sqrt{L^{2}}}
\left(
	\begin{array}{c}
		ie^{-i\phi_{\bm{k}}}\cos \frac{\theta_{\bm{k}}}{2} \\
		\sin \frac{\theta_{\bm{k}}}{2}
	\end{array}
\right),
\psi_{-} = \frac{e^{i\bm{k}\cdot \bm{r}}}{\sqrt{L^{2}}}
\left(
	\begin{array}{c}
		\sin \frac{\theta_{\bm{k}}}{2} \\
		ie^{i \phi_{\bm{k}}}\cos \frac{\theta_{\bm{k}}}{2}
	\end{array}
\right) 
\end{align}
\begin{align}
\epsilon_{\bm{k},s} = s \sqrt{v^{2}k^{2}+\alpha^{2}k^{6}\cos^{2}3\phi_{\bm{k}}},
\end{align}
with $\phi_{\bm{k}}=\tan^{-1} (k_{y}/k_{x})$ and $\tan \theta_{\bm{k}} = {k_{c}^{2}}/({k^{2}\cos 3\phi_{\bm{k}}})$, where $k_{c}=\sqrt{v/\alpha}$. It is apparent  that each eigenstate has nonzero expectation value of $\sigma_{z}$ as
\begin{align}
\langle \bm{k}s |\sigma_{z}|\bm{k}s \rangle = s \cos \theta_{\bm{k}} \not\equiv 0,
\end{align}
which is schematically depicted in Fig.~\ref{fig:warping_spins}.

To investigate the current induced spin polarization, we expand the spin density $\langle \sigma_{z} \rangle$ in terms of the current $\langle j_{i}\rangle \ (i=x,y)$ with coefficients $A_{ij}$:
\begin{align}
\langle \sigma_{z}\rangle = \sum_{i,j}A_{i,j}\langle j_{x}\rangle^{i}\langle j_{y}\rangle^{j}
\end{align}
where the expression $A_{ij}$ is restricted by the symmetry of the system.

The Hamiltonian (Eq.~(\ref{warping_hamiltonian})) possesses the time-reversal symmetry ${\cal T}$, the threefold rotational symmetry around the $z$-axis $C_{3z}$, and the mirror symmetry with respect to the $yz$-plane, $M_{yz}$. The mirror symmetry $M_{yz}$ gives $\langle j_{x}\rangle \to - \langle j_{x}\rangle$, $\langle j_{y}\rangle \to \langle j_{y}\rangle$ and $\langle \sigma_{z} \rangle \to - \langle \sigma_{z} \rangle$, and therefore $\langle \sigma_{z} \rangle$ is an odd function about $\langle j_{x} \rangle$ and $A_{ij}$ is allowed to be nonzero only for $i={\rm odd}$. In addition, because of the time-reversal symmetry, $\langle \sigma_{z}\rangle$ is an even function of $\langle j_{y}\rangle$ and $A_{ij}$ is allowed to be nonzero only for $j={\rm even}$.
Furthermore, $\langle \sigma_{z} \rangle$ is invariant under the three-fold rotation $C_{3z}$: $(\langle j_{x}\rangle,\, \langle j_{y}\rangle ) \to (-\frac{1}{2}\langle j_{x}\rangle - \frac{\sqrt{3}}{2}\langle j_{y}\rangle,\, \frac{\sqrt{3}}{2}\langle j_{x}\rangle -\frac{1}{2}\langle j_{y}\rangle)$, which gives a restriction on the coefficients $A_{ij}$. For example, $i=1,\ j=0$ term can not be invariant under the $2\pi /3$ rotation, but the $i=1,\ j=2$ and $i=3,\ j=0$ terms remain the same under this rotation if $A_{30}=-A_{12}/3$. From these symmetry considerations, the expansion of the spin polarization with respect to the current in the lowest order yields
\begin{align}
\langle \sigma_{z}\rangle \propto \langle j_{x}\rangle^{3} - 3\langle j_{x}\rangle \langle j_{y}\rangle^{2},
\end{align}
which is cubic of the current. When the external field is small, the current is linearly dependent on the field, and the spin polarization can be written as $\langle \sigma_{z}\rangle \propto E_{x}^{3} - 3E_{x} E_{y}^{2}$. The spin polarization depends not only on the magnitude but also the direction of the electric field. The spin polarization is largest when the field is along the $x$ axis, and zero when the field is tilted by $\pi/6$ from the $x$ axis. 
\section{CONCLUSION}
In this paper, we have investigated the response of the surface of TI to the electric fields. We have used the Keldysh formalism to calculate the spin polarization of the surface of TI in a systematic fashion. In the linear response regime, we have found that dc current can induce the spin polarization perpendicular to the current due to the strong spin-orbit coupling. Also, we have estimated the magnitude of the spin polarization. We have further investigated the inverse Faraday effect of the TI based on the second order perturbation, and we have found that  the spin polarization emerges by circularly polarized light illumination, which is proportional to square of the lifetime $\tau$. 
We also investigated effects of the warping term and found that there is a spin polarization perpendicular to the surface of TIs which has the three fold rotational symmetry and is cubic with respect to the current.

The authors thank K. Taguchi for helpful discussion.
This work is partly supported by Grant-in-Aids from MEXT, Japan (No. 21000004 and 22540327), and by the Global Center of Excellence Program by MEXT, Japan through the "Nanoscience and Quantum Physics" Project of the Tokyo Institute of Technology 
and also by Grant-in-Aid for Young Scientists (B) (No. 23740236) and the "Topological Quantum Phenomena" (No. 23103505) Grant-in Aid for Scientific Research on Innovative Areas from the Ministry of Education, Culture, Sports, Science and Technology (MEXT) of Japan.

\appendix
\section{Calculation of the Green's function in Eq.(5)}
Here, we show how the $g^{\rm r(a)}Vg^{\rm r(a)}$ terms vanish in Eq.~(\ref{first_order_perturbation}).  
The key is that $V$ is written in the form of current operator. Hence, the vertex function can be written in the form of the derivative of the Hamiltonian in terms of $k_{x}$:
\begin{align}
\tilde{V}_{\Omega}=\frac{ieE_{x}}{\Omega} \frac{\partial H_{0}}{\partial k_{x}},
\end{align}
and hence we have 
\begin{align}
g^{\rm r(a)}\tilde{V}_{\Omega}g^{\rm r(a)} = \frac{ieE_{x}}{\Omega}\frac{\partial g^{\rm r(a)}}{\partial k_{x}}.
\end{align}

Considering the integration of these terms about $k$, we obtain
\begin{align}
\int \frac{\mathrm{d}^{2}k}{4\pi^{2}} g^{\rm r(a)}\tilde{V}_{\Omega}g^{\rm r(a)}
& \propto \int \frac{\mathrm{d}^{2}k}{4\pi^{2}} \frac{\partial g^{\rm r(a)}}{\partial k_{x}} \nonumber \\
& = \frac{1}{4\pi^{2}} \int \mathrm{d}k_{y}\bigl. g^{\rm r(a)}\Bigr|_{k_{x}=-\infty}^{k_{x}=+\infty},
\end{align}
where the integrand is zero from the periodicity of the system. Therefore, the integration becomes zero after all.

\section{Calculation of the Green's function in Eq.(12)}
Here, we show details of the calculation of the Green's function in Eq.(12). In this case, $\tilde{V}_{\Omega}^{(\dagger)}$ is proportional to the derivative of the Hamiltonian with respect to $k_{\pm}$:
\begin{align}
\tilde{V}_{\Omega} = \frac{2ie{\cal E}}{\Omega} \frac{\partial H_{0}}{\partial k_{+}},\ \tilde{V}_{\Omega}^{\dagger} = -\frac{2ie{\cal E}}{\Omega} \frac{\partial H_{0}}{\partial k_{-}}.
\end{align}
Therefore, we have
\begin{align}
g^{\rm r(a)}\tilde{V}_{\Omega}g^{\rm r(a)} = \frac{2ie{\cal E}}{\Omega} \frac{\partial g^{\rm r(a)}}{\partial k_{+}},\
g^{\rm r(a)}\tilde{V}_{\Omega}^{\dagger}g^{\rm r(a)} = -\frac{2ie{\cal E}}{\Omega} \frac{\partial g^{\rm r(a)}}{\partial k_{-}}.
\end{align}

The terms independent of $\Omega$ and composed of only $g^{\rm r(a)}$ in the Eq.~(\ref{second_order_perturbation}) are shown to be vanish as follows
\begin{widetext}
\begin{align}
\int \frac{\mathrm{d}^{2}k}{4\pi^{2}}\, \left(g^{\rm r}\tilde{V}_{\Omega}g^{\rm r}\tilde{V}_{\Omega}^{\dagger}g^{\rm r} + g^{\rm r}\tilde{V}_{\Omega}g^{\rm r}\tilde{V}_{\Omega}g^{\rm r} - ({\rm r}\leftrightarrow {\rm a}) \right)
& \propto \int \frac{\mathrm{d}^{2}k}{4\pi^{2}} \, \left[ \frac{\partial g^{\rm r}}{\partial k_{+}}(g^{\rm r})^{-1}\frac{\partial g^{\rm r}}{\partial k_{-}}+ \frac{\partial g^{\rm r}}{\partial k_{-}}(g^{\rm r})^{-1}\frac{\partial g^{\rm r}}{\partial k_{+}}  - ({\rm r}\leftrightarrow {\rm a}) \right] \nonumber \\
& = \int \frac{\mathrm{d}^{2}k}{4\pi^{2}} \, \left[ g^{\rm r}\frac{\partial (g^{\rm r})^{-1}}{\partial k_{+}\partial k_{-}}g^{r} + g^{\rm r} \frac{\partial (g^{\rm r})^{-1}}{\partial k_{-}} \frac{\partial g^{\rm r}}{\partial k_{+}} +\frac{\partial g^{\rm r}}{\partial k_{-}}(g^{\rm r})^{-1}\frac{\partial g^{\rm r}}{\partial k_{+}} - ({\rm r}\leftrightarrow {\rm a}) \right] \nonumber \\
& = 0, \label{Omega_zeroth}
\end{align}
\end{widetext}
where we used the partial integration and made use of the relation $\partial (g^{\rm r})^{-1}g^{\rm r}/\partial k_{\pm} =0$.
Note also that $(g_{\omega}^{\rm r(a)})^{-1} = \omega - H_{0} \pm i\eta_{\omega}$ and the Hamiltonian $H_{0}$ is linear about $k$. 

Next, we show that the terms linear in $\Omega$ which consist of only $g^{\rm r(a)}$ vanish as follows:
\begin{widetext}
\begin{align}
\int \frac{\mathrm{d}^{2}k}{4\pi}\, f(\omega) \Omega \left[ g^{\rm r} \tilde{V}_{\Omega} \frac{\partial g^{\rm r}}{\partial \omega}\tilde{V}_{\Omega}^{\dagger}g^{\rm r}-g^{\rm r} \tilde{V}_{\Omega}^{\dagger} \frac{\partial g^{\rm r}}{\partial \omega} \tilde{V}_{\Omega} g^{\rm r} -({\rm r}\leftrightarrow {\rm a}) \right]
& = - f(\omega) \Omega \int \frac{\mathrm{d}^{2}k}{4\pi^{2}}\, \left[ g^{\rm r} \tilde{V}_{\Omega} g^{\rm r}g^{\rm r}\tilde{V}_{\Omega}^{\dagger}g^{\rm r}-g^{\rm r} \tilde{V}_{\Omega}^{\dagger} g^{\rm r}g^{\rm r} \tilde{V}_{\Omega} g^{\rm r}-({\rm r}\leftrightarrow {\rm a}) \right] \nonumber \\
& \propto - f(\omega) \Omega \int \frac{\mathrm{d}^{2}k}{4\pi^{2}}\, \left[ \frac{\partial g^{\rm r}}{\partial k_{+}} \frac{\partial g^{\rm r}}{\partial k_{-}} - \frac{\partial g^{\rm r}}{\partial k_{-}}\frac{\partial g^{\rm r}}{\partial k_{+}} - ({\rm r}\leftrightarrow {\rm a}) \right] \nonumber \\
& = 0,
\end{align}
\end{widetext}
where we again used the partial integration. 
As for those terms which include the derivative of the Fermi distribution function, we can take a similar procedure to show that they vanish:
\begin{widetext}
\begin{align}
& \int \frac{\mathrm{d}^{2}k}{4\pi^{2}} \, \Omega f'(\omega) \left[g^{\rm r}\tilde{V}_{\Omega}(g^{\rm a}-g^{\rm r})\tilde{V}_{\Omega}^{\dagger}g^{\rm a}-g^{\rm r}\tilde{V}_{\Omega}^{\dagger}(g^{\rm a}-g^{\rm r})\tilde{V}_{\Omega}g^{\rm a}\right] \nonumber \\
& \hspace{2em} \propto \Omega f'(\omega) \int \frac{\mathrm{d}^{2}k}{4\pi^{2}} \, \left[ g^{\rm r}\frac{\partial H_{0}}{\partial k_{+}}\frac{\partial g^{\rm a}}{\partial k_{-}}-\frac{\partial g^{\rm r}}{\partial k_{+}}\frac{\partial H_{0}}{\partial k_{-}}g^{\rm a}-g^{\rm r}\frac{\partial H_{0}}{\partial k_{-}}\frac{\partial g^{\rm a}}{\partial k_{+}}+\frac{\partial g^{\rm r}}{\partial k_{-}}\frac{\partial H_{0}}{\partial k_{+}}g^{\rm a} \right] \nonumber \\
& \hspace{2em} = \Omega f'(\omega) \int \frac{\mathrm{d}^{2}k}{4\pi^{2}} \, \left[
-\frac{\partial }{\partial k_{-}}\left(g^{\rm r}\frac{\partial H_{0}}{\partial k_{+}}\right)g^{\rm a}+g^{\rm r} \frac{\partial}{\partial k_{+}}\left( \frac{\partial H_{0}}{\partial k_{-}}g^{\rm a}\right) -g^{\rm r}\frac{\partial H_{0}}{\partial k_{-}}\frac{\partial g^{\rm a}}{\partial k_{+}} + \frac{\partial g^{\rm r}}{\partial k_{-}}\frac{\partial H_{0}}{\partial k_{+}} g^{\rm a}\right] \nonumber \\
& \hspace{2em} = - \Omega f'(\omega) \int \frac{\mathrm{d}^{2}k}{4\pi^{2}} \ \left( g^{\rm r}\frac{\partial^{2}H_{0}}{\partial k_{-}\partial k_{+}}g^{\rm a} - g^{\rm r}\frac{\partial^{2}H_{0}}{\partial k_{+}\partial k_{-}} g^{\rm a}\right) \nonumber \\
& \hspace{2em} = 0,
\end{align}
\end{widetext}
since $\partial^{2}H_{0}/\partial k_{+}\partial k_{-} =0$.

The terms proportional to $\Omega^{0}$ or $\Omega^{1}$ have to be zero because otherwise some physical quantities, such as electron density $\propto {\rm Tr}G^{<}(\bm{r},t,\bm{r},t)$, diverge in the low frequency limit $\Omega \searrow 0$. 

Regarding the terms proportional to $\Omega^{2}$, we find that these vanish after multiplying the corresponding term by $\sigma_{z}$ and taking trace:
\begin{widetext}
\begin{align}
&{\rm Tr}\, \sigma_{z} \Omega^{2}
	\Biggl[ \frac{f(\omega)}{2}
		\left[
			\left\{
				\left( g_{\omega}^{\rm r}\tilde{V}_{\Omega}\frac{\partial^{2} g^{\rm r}_{\omega}}{\partial \omega^{2}}\tilde{V}_{\Omega}^{\dagger}g_{\omega}^{\rm a-r} + g_{\omega}^{\rm r}\tilde{V}_{\Omega}^{\dagger}\frac{\partial^{2} g^{\rm r}_{\omega}}{\partial \omega^{2}}\tilde{V}_{\Omega}g_{\omega}^{\rm a-r}
				\right) -{\rm h.c.}
			\right\} + g_{\omega}^{\rm r}\tilde{V}_{\Omega}\frac{\partial^{2} g^{\rm a-r}_{\omega}}{\partial \omega^{2}}\tilde{V}_{\Omega}^{\dagger}g_{\omega}^{\rm a} + g_{\omega}^{\rm r}\tilde{V}_{\Omega}^{\dagger} \frac{\partial^{2} g^{\rm a-r}_{\omega}}{\partial \omega^{2}}\tilde{V}_{\Omega}g_{\omega}^{\rm a}
		\right]
	\Biggr. \hspace{4em} \nonumber \\
& \hspace{28em} \Biggl. + f'(\omega) \left[ g_{\omega}^{\rm r}\tilde{V}_{\Omega}\frac{\partial g^{\rm a-r}_{\omega}}{\partial \omega}\tilde{V}_{\Omega}^{\dagger}g_{\omega}^{\rm a} 
- g_{\omega}^{\rm r}\tilde{V}_{\Omega}^{\dagger}\frac{\partial g^{\rm a-r}_{\omega}}{\partial \omega}\tilde{V}_{\Omega}g_{\omega}^{\rm a} \biggl. \right] \Biggr] = 0,
\end{align}
\end{widetext}
with $g^{\rm a-r}=g^{\rm a}-g^{\rm r}$. 

\section{Comparison to the Rashba system}
The previous studies on the Rashba systems~\cite{Rashba}  have shown that the dc current-induced spin accumulation occurs in the Rashba system~\cite{Ede90,Ino03,Sinova}. In this Appendix, we introduce the calculation of longitudinal conductivity and the spin accumulation in the Rashba system. The calculation methods used below are presented in Ref.~[\onlinecite{Ino03}].

The model Hamiltonian for the Rashba system is
\begin{align}
H_{0}^{\rm Rashba}=\frac{k^{2}}{2m}+\alpha (\bm{\sigma}\times \bm{k})_{z} \label{rashba_hamiltonian},
\end{align}
where the $\alpha$ is the magnitude of the SOC. 
The eigenvalues are given by $E_{\bm{k}s}=k^{2}/2m + s \alpha k$, where $k=\sqrt{k_{x}^{2}+k_{y}^{2}}$ and $s= \pm$, and the corresponding eigenstates are $\phi_{s}(\bm{r})=e^{i\bm{k}\cdot \bm{r}}/{\sqrt{2L^{2}}}(1,\, -isk_{+}/k)$.
It is notable that the Rashba Hamiltonian has the diagonal component in its Hamiltonian in contrast to the case of the TI. The current operator is $j_{x}=e(bk_{x}-\alpha \sigma_{y})$ and $j_{y}=e(bk_{y}+\alpha \sigma_{x})$
with $b=1/m$.
The unperturbed Green's function is $g_{0}=[z-H_{0}^{\rm Rashba}]^{-1}$.
Taking into account spin-independent short-range impurity potential $V(\bm{r})= V\sum_{i=1}^{N_{imp}} \delta (\bm{r}-\bm{R}_{i})$,  the self-energy can be derived. The modification can be made via the self-energy $\eta$ in a similar manner to the case of the TI. We here employ the Born approximation, and the perturbed Green's function reads $g^{\rm r(a)} = [z-H_{0}^{\rm Rashba} \pm i \eta]^{-1}$, where $\eta =mn_{\rm imp}V_{\rm imp}^{2}/2$ is the magnitude of the self-energy.

Before using the Green's function obtained above to calculate the longitudinal conductivity, the vertex correction has to be considered. With the iterative method used in the case of the TI, we obtain the modified current operator $\tilde{j}_{x}$, which is obtained by doubling the $\alpha$ in $j_{x}$. The conductivity is then calculated via the Kubo formula as
\begin{align}
\sigma_{xx} = \frac{1}{2\pi L^{2}} {\rm Tr} j_{x}g^{\rm r}\tilde{j}_{x}g^{\rm a} = \frac{2e^{2}n_{0}\tau}{m}+\frac{e^{2}\alpha^{2}}{\pi n_{\rm imp}V_{\rm imp}^{2}} \label{rashba_conductivity}
\end{align}
with the $\tau$ being the relaxation time, and $n_{0}$ is the local  electron density. The contribution of the first term of Eq.~(\ref{rashba_conductivity}) is the same as that of the Drude model, and we can see that the second term is the same as the conductivity of the surface state of the TI. The calculation of the spin accumulation can be done in a similar way, which yields
\begin{align}
\langle \sigma_{y} \rangle = \frac{e\alpha}{2\pi n_{\rm imp}V_{\rm imp}^{2}}E_{x},
\end{align}
which is half the spin accumulation on the surface of the TI.


\section{Calculation by the Kubo formula}
In the previous sections, we adopted the Keldysh formalism, which is convenient  to deal with higher order perturbations. In this section, we perform the calculation of the spin polarization and the longitudinal conductivity of the TI via the Kubo formula~\cite{Mahan}.

\begin{figure}[htbp]
	\begin{center}
		\includegraphics[width = 7cm,keepaspectratio,clip]{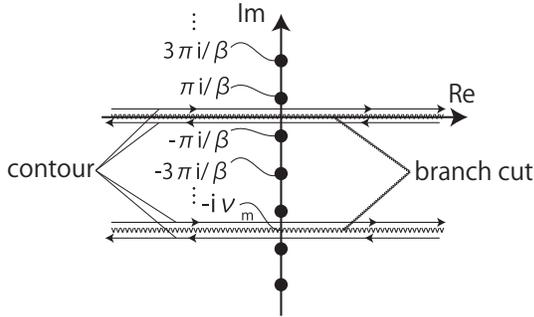}
		\caption{The contour of the integration. The contour is deformed to avoid the branch cuts on the complex plane.}
		\label{fig:contour}
	\end{center}
\end{figure}

We perform the calculations by means of the Matsubara Green's function ${\cal G}$.
To calculate the response functions of spin polarization $\phi_{i}\ (i=x,y,z)$, which are defined as $\langle \sigma_{i} \rangle = \phi_{i} E_{x}$, we first compute the spin-current correlation function
\begin{equation}
	Q_{i}(i\nu_{m} )= \frac{1}{L^{2}\beta}\sum_{\bm{k}n}\mathrm{Tr}\, \sigma_{i} {\cal G}_{\bm{k}}(i\omega_{n} + i\nu_{m}) \tilde{j}_{x} {\cal G}_{\bm{k}}(i\omega_{n}) , \label{kubo_matsubara}
\end{equation}
where $\beta=(k_{\rm B}T)^{-1}$, $\nu_{n}=n\pi /\beta \ (n=0,\, \pm 2,\, \pm4,\cdots )$ and $\omega_{n}= n \pi /\beta \ (n=\pm 1,\, \pm 3,\, \pm5,\cdots )$. Subsequently, we make the analytic continuation $i\nu_{m} \to \omega+i\delta$ and take the limit $\lim_{\omega \to 0} Q_{i}(\omega)/i\omega$ to obtain the response functions.
The summation over $n$ on the right hand side of Eq.~(\ref{kubo_matsubara}) is done by replacing the summation with the contour integral after multiplying that by the Fermi distribution function. 
With the aid of the residual theorem, we can represent the correlation function as
\begin{align}	
	\frac{-1}{L^{2}}\int_{C}\frac{\mathrm{d}z}{2\pi i}\frac{1}{e^{\beta z}+1}\sum_{\bm{k}}\mathrm{Tr} \, \sigma_{i}{\cal G}_{\bm{k}}(z+ i\nu_{m})\tilde{j}_{x}{\cal G}_{\bm{k}}(z),
\end{align}
where we took the integral contour ${\cal C}$ as seen in Fig.~\ref{fig:contour}.
The integration can be further rewritten as 
\begin{widetext}
	\begin{align}
	Q_{i}(i\nu_{m}) &= \frac{1}{L^{2}}\int_{-\infty}^{\infty} \frac{\mathrm{d}\epsilon}{2\pi}f(\epsilon ) \sum_{\bm{k}}\mathrm{Tr}\Bigl[ \sigma_{i} {\cal G}_{\bm{k}}(\epsilon +i\nu_{m})\tilde{j}_{x}A_{\bm{k}}(\epsilon ) + \sigma_{i} A_{\bm{k}}(\epsilon )\tilde{j}_{x} {\cal G}_{\bm{k}}(\epsilon - i\nu_{m}) \Bigr],
	\end{align}
\end{widetext}
where $f(z) = (e^{\beta z}+1)^{-1}$ is the Fermi distribution function and $-i A_{\bm{k}}(\epsilon)\equiv {\cal G}_{\bm{k}}(\epsilon +i\delta )-{\cal G}_{\bm{k}}(\epsilon -i\delta )$.

By the analytic continuation $i\nu_{m} \to \omega + i\delta$ and $\mathcal{G}(\omega \pm i\delta)=g^{\rm r(a)}_{\omega}$, the correlation function $Q_{i}$ yields
\begin{align}
	Q_{i}(\omega +i\delta)
 & = -\frac{1}{L^{2}}  \int \frac{\mathrm{d}\epsilon}{2\pi i}\{-f(\epsilon)+f(\epsilon+ \omega )\} \nonumber \\
 & \hspace{5em} \times \sum_{\bm{k}} \mathrm{Tr}\, \sigma_{i}g^{\rm r}_{\bm{k},\epsilon+\omega} \tilde{j}_{x}g^{\rm a}_{\bm{k},\epsilon}
\label{rashba_retarded_green}.
\end{align}
We here consider the zero temperature limit where the Fermi distribution function becomes the step function, and hence
$ \{ -f(\epsilon) + f(\epsilon+ \omega )\} / \omega \to - \delta (\epsilon) \ (\omega \to 0)$.
Then we obtain the response functions $\phi_{i}$ as:
\begin{align}
	\phi_{i}=& \frac{1}{2\pi L^{2}} \sum_{\bm{k}}\mathrm{Tr}\, \sigma_{i}g^{\rm r}_{\bm{k},\epsilon_{\rm f}}\tilde{j}_{x}g^{\rm a}_{\bm{k},\epsilon_{\rm F}},
\end{align}
which gives $\phi_{x}=\phi_{z} = 0$ and
\begin{align}
\phi_{y} = \frac{ev}{\pi n_{\rm imp}V_{\rm imp}^{2}}.
\end{align}
These results are consistent with the results by the Keldysh formalism as expected.

Similarly, we obtain the longitudinal conductivity by substituting $\sigma_{i}$ with $j_{x}$ in Eq.~(\ref{kubo_matsubara}). This leads to
\begin{align}
	\sigma_{xx} = \frac{1}{2\pi L^{2}} \sum_{\bm{k}}{\rm Tr} \, j_{x} g^{\rm r}_{\bm{k},\epsilon_{\rm F}} \tilde{j}_{x} g^{\rm a}_{\bm{k},\epsilon_{\rm F}},
\label{dc_conductivity_kubo}
\end{align}
 and finally we obtain
\begin{align}
\sigma_{xx} = \frac{e^{2} v^{2}}{\pi n_{\rm imp}V_{\rm imp}^{2}},
\end{align}
which is in accordance with the expression obtained by the Keldysh formalism.


\end{document}